\documentstyle[aps,prl]{revtex}

\renewcommand{\theequation}{\thesection.\arabic{equation}}
\newcounter{hran} \renewcommand{\thehran}{\thesection.\arabic{hran}}

\def\bmini{\setcounter{hran}{\value{equation}}
  \refstepcounter{hran}\setcounter{equation}{0}
  \renewcommand{\theequation}{\thehran\alph{equation}}\begin{eqnarray}}

\def\bminiG#1{\setcounter{hran}{\value{equation}}
\refstepcounter{hran}\setcounter{equation}{-1}
\renewcommand{\theequation}{\thehran\alph{equation}}
\refstepcounter{equation}\label{#1}\begin{eqnarray}}

\renewcommand{\theequation}{\thesection.\arabic{equation}}

\def\bmini{\setcounter{hran}{\value{equation}}
  \refstepcounter{hran}\setcounter{equation}{0}
  \renewcommand{\theequation}{\thehran\alph{equation}}\begin{eqnarray}}

\def\bminiG#1{\setcounter{hran}{\value{equation}}
\refstepcounter{hran}\setcounter{equation}{-1}
\renewcommand{\theequation}{\thehran\alph{equation}}
\refstepcounter{equation}\label{#1}\begin{eqnarray}}

\def\emini{\end{eqnarray}\relax\setcounter{equation}
{\value{hran}}\renewcommand{\theequation}{\thesection.\arabic{equation}}}

\newcommand{\bal}{\begin{array}{ll}} 
\newcommand{\eal}{\end{array}}

\def\1{{\rm 1 \kern -.10cm I \kern .14cm}} \def\R{{\rm R \kern -.28cm I
\kern .19cm}}
\def\emini{\end{eqnarray}\relax\setcounter{equation
}{\value{hran}}\renewcommand{\theequation}{\thesection.\arabic{equation}}}

\newcommand{\be}{\begin{equation}}
\newcommand{\ee}{\end{equation}}
\def\bea{\begin{eqnarray}}
\def\eea{\end{eqnarray}}

\newcommand{\barre}[1]{%
	\setbox1=\hbox{$#1$} \dimen2=\ht1 \dimen3=\dp1 \dimen4=\wd1
	\setbox2=\hbox{\sl /}
	\dimen1=\wd1 \advance\dimen1 by -\wd2 \divide\dimen1 by 2
	\advance\dimen1 by \wd2 \advance\dimen1 by 0.4pt
	\setbox3=\hbox to \wd1{\hss \box1 \kern -\dimen1 \box2\hss}
	\ht3=\dimen2 \dp3=\dimen3 \wd3=\dimen4
	\box3
	}

\def\1{{\rm 1 \kern -.10cm I \kern .14cm}} \def\R{{\rm R \kern -.28cm I
\kern .19cm}}
\begin{document}
\begin{titlepage}
\begin{flushright}    UFIFT-HEP-98-14 \\ 
\end{flushright}
\vskip 1cm
\centerline{\LARGE{\bf {Family Symmetry and Neutrino Mixing}}}
\vskip 1.5cm
\centerline{\bf John K. Elwood$^2$ \footnote{e-mail address 
jelwood@phys.ufl.edu}, Nikolaos Irges$^1$ \footnote{e-mail address
irges@phys.ufl.edu}
and Pierre Ramond$^1$ \footnote{e-mail address ramond@phys.ufl.edu}} 

 
\vskip .7cm
\centerline{\em  $^1$Institute for Fundamental Theory,}
\centerline{\em Department of Physics, University of Florida}
\centerline{\em Gainesville FL 32611, USA}
\vskip .2cm
\centerline{\em $^2$Department of Physics, Kent State
University,}
\centerline{\em Kent, Ohio 44242, USA} 
\vskip 1.5cm
\centerline{\bf {Abstract}}
\vskip .5cm
The observed quark hierarchies suggest a simple family 
symmetry. Generalized to leptons through grand-unified 
quantum numbers, it produces a neutrino mixing matrix with 
order-one $\nu_\mu-\nu_\tau$ mixing, 
and order-$\lambda^3$ $\nu_e-\nu_\mu$ and $\nu_e-\nu_\tau$
mixings. The intrafamily hierarchy and 
observed neutrino mass differences together require this symmetry
to be anomalous, suggesting through the Green-Schwarz mechanism a
string or M-theory origin for the symmetry.
 \vfill
\begin{flushleft}
June 1998 \\
\end{flushleft}
\end{titlepage}
\section{Introduction}
The Yukawa sector of the Standard Model is clearly its most poorly
understood aspect, with a number of undetermined masses and mixings.
Its extension  to include low energy supersymmetry, however,
brings about several conceptual simplifications including the unification 
of its gauge couplings and the equality of the  bottom quark and 
tau lepton Yukawa couplings  at a scale $M_{GUT}\sim 10^{16}\;$ GeV.
Further regularities in the Yukawa sector are best appreciated
by using the Wolfenstein 
parametrization of the CKM matrix \cite{wolf},
\be \pmatrix{1&\lambda & \lambda ^3\cr
             \lambda &1&\lambda ^2\cr 
             \lambda ^3&\lambda ^2&1} \ \ \ \ , \ee
in terms of the Cabibbo angle $\lambda$, with all prefactors of order one.
Extended to the  quark and charged lepton mass ratios, it reveals  both
an {\bf interfamily} hierarchy
\be {m_{u}\over m_t}\sim \lambda ^8\;\;\;{m_c\over m_t}\sim 
\lambda ^4\;\;\; ;
\;\;\; {m_d\over m_b}\sim \lambda ^4\;\;\;{m_s\over m_b}\sim
\lambda^2
\ \ ; {m_e\over m_\tau}\sim \lambda ^4\;\;\;{m_\mu\over m_\tau}\sim
\lambda^2\ \ ,\ee
and an {\bf intrafamily} hierarchy
\be {m_b\over m_t}\sim \lambda ^3\;\;\;{m_b\over m_{\tau}}\sim 1 \ \ ,\ee
valid at  the
scale $M\sim M_{GUT}$.
Such relations become apparent if low energy supersymmetry
is realized.

Below we argue for a simple explanation: the interfamily 
hierarchy points to the 
existence of at least one 
$U(1)$ family symmetry beyond the Standard Model. In addition, the 
intrafamily hierarchy, together with the recent neutrino 
data\cite{superk}, requires one of these 
to be anomalous\cite{IR,BR}.  

In our approach, the Standard Model is an effective field theory  with
cut-off $M\sim M_{GUT}$. Standard Model invariants may not appear in
the superpotential unless they also have zero charge with respect to
any additional symmetries; in this way, invariants with non-zero
charges will be present only as higher dimension operators, leading to
the suppression of their couplings. Thus,  the exponents that appear
in the hierarchies can be directly related to the charges of the
accompanying SM invariants\cite{FN,RRR}. This letter addresses only
the origin of the Cabibbo suppressions of the Yukawa couplings, not
the supersymmetry nor the electroweak breaking mechanisms.
\section{Interfamily Hierarchy}

\par To study the interfamily hierarchy, it suffices to consider the
relative strengths of the Yukawa couplings within each charge sector
separately.
 In the down quark sector, the observed hierarchy allows us to reconstruct 
part of the $3\times 3$ Yukawa matrix that labels the strength of the 
operators
${\bf Q}_i{\bf \overline d}_jH_d$. Assuming the matrices that
diagonalize the Yukawa couplings are, like the CKM matrix, nearly
diagonal, the mass ratios determine its
$(11)$ 
and $(22)$
elements to be of order $\lambda ^4$ and $\lambda ^2$, respectively, 
as compared to 
the $(33)$ element, which we take to be of order of 
one.\footnote{Since here we 
are interested only in the 
interfamily hierarchy, we neglect for the time being 
an overall factor that may multiply these elements.} 
Our first assumption is that both matrices that make up the CKM matrix have the same Cabibbo
structure (no alignment\footnote{This means our
model must rely on a supersymmetry-breaking mechanism that generates no
large FCNC effects, whose discussion is beyond the scope of this letter.}). Then,  the upper off-diagonal elements are determined by the CKM matrix,
yielding the exponents

\hskip 2cm
\begin{center}
\begin{tabular}{|c|c|c|c|}
\hline          
$ $ & ${\rm {\bf \overline d}_1}$ & ${\bf \overline d}_2$  & 
${\bf \overline d}_3$  \\ \hline \hline   

${\bf Q}_1$ & $4$ & $3$ & $3$ \\

${\bf Q}_2$ & $ $ & $2$ & $2$ \\          
                                                   
${\bf Q}_3$ & $ $ & $ $ & $0$ \\ \hline
\end{tabular} \ \ \ \ \ .
\end{center}
\vskip 0.3cm 
The orders of magnitude of elements below the 
diagonal are not determined by
phenomenology. Our second  assumption is that the exponents are determined by
the charges of the operators under a family symmetry $Y_F$. 
The exponents in the $(ij)$ entry, for
example, 
would be proportional to $(x_{{\bf Q}_i}+x_{{\bf \overline d}_j}+x_{H_d})$.
The charge of $H_d$ does not affect the interfamily patterns. 
This origin for the exponents 
implies sum rules between them $(n_{ij}+n_{ji}=n_{ii}+n_{jj})$~
\cite{BR},
leading to the full matrix:

\hskip 2cm
\begin{center}
\begin{tabular}{|c|c|c|c|}
\hline          
$ $ & ${\rm {\bf \overline d}_1}$ & ${\bf \overline d}_2$  & 
${\bf \overline d}_3$  \\ \hline \hline   

${\bf Q}_1$ & $4$ & $3$ & $3$ \\

${\bf Q}_2$ & $3$ & $2$ & $2$ \\  
                                                           
${\bf Q}_3$ & $1$ & $0$ & $0$ \\ \hline
\end{tabular} \ \ \ \ \ .
\end{center} 
\vskip 0.3cm 

\noindent When an operator's charge prevents it from appearing 
in the superpotential as a holomorphic invariant \cite{LENS}, 
we say that the operator develops
a ``supersymmetric zero''. Unless otherwise specified, we will assume
throughout this paper, as we just did, that there are no 
supersymmetric zeros. From 
the above matrix, we can read off the charges of the 
${\bf \overline d}_i$ and ${\bf Q}_i$:
\be ({\bf \overline d}_1,{\bf \overline d}_2,{\bf \overline d}_3)=
-({1 \over 3})(2,-1,-1) \ \ \ , \ee
\be ({\bf Q}_1,{\bf Q}_2,{\bf Q}_3)=
-({1 \over 3})(4,1,-5)\ .\ee
Since we are discussing only the interfamily hierarchies, we need 
consider only the family-traceless  charges.  Suggestively, we can 
express these in terms of baryon
number as
\be ({\bf \overline d}_1,{\bf \overline d}_2,{\bf \overline d}_3)=
B(2,-1,-1) \ \ \ , \ee 
\be ({\bf Q}_1,{\bf Q}_2,{\bf Q}_3)=
B(2,-1,-1)-2(1,0,-1).\ee
In the charge $2/3$ sector,
the $(11)$ and $(22)$ elements have exponents $8$ and $4$ 
respectively, and using the CKM matrix, we obtain:

\hskip 2cm
\begin{center}
\begin{tabular}{|c|c|c|c|}
\hline          
$ $ & ${\rm {\bf \overline u}_1}$ & ${\bf \overline u}_2$  & 
${\bf \overline u}_3$  \\ \hline \hline   

${\bf Q}_1$ & $8$ & $5$ & $3$ \\

${\bf Q}_2$ & $7$ & $4$ & $2$ \\
                                                            
${\bf Q}_3$ & $5$ & $2$ & $0$ \\ \hline
\end{tabular} \ \ \ \ ,
\end{center}
\vskip 0.3cm 

\noindent where the entries below the diagonal are determined  
by sum rules. We arrive of course at exactly the same charges 
for the ${\bf Q}_i$,
but find also the ${\bf \overline u}_i$ charge assignment 
\be ({\bf \overline u}_1,{\bf \overline u}_2,{\bf \overline u}_3)=
B(2,-1,-1)-2(1,0,-1).\ee
Thus, the family dependent symmetry acting upon the quark sector may
be written
\be Y_F=B(2,-1,-1)-2\eta(1,0,-1)\label{eq:yf} ,\ee
where $\eta=1$ for both ${\bf Q}$ and ${\bf \overline u}$ and 
$\eta=0$ for ${\bf \overline d}$.  That both 
${\bf Q}$ and ${\bf \overline u}$  possess the same $\eta$ charge
is reminiscent of the $SU(5)$ charge patterns, 
where  the chiral fermions are split 
into $\overline {\bf{5}}=(L,{\bf {\overline d}})$ and ${\bf{10}}=({\bf
Q},{\bf {\overline u}},\overline
e)$. Our third assumption uses these Grand Unified patterns to
assign family charges to the leptons: we flesh out the multiplets by assigning the $\overline{e}$
singlet a value $\eta=1$ and the $L$ doublet $\eta=0$, and 
generalize the factor $B$ appearing in
Eq.~(\ref{eq:yf}) to its $SO(10)$ analog $(B-L)$.  
Note that $\eta$, on the other hand, is outside of $SO(10)$.
The quark and lepton charges may
then be succinctly written:
\be
Y_F=(B-L) (2, -1, -,1) - 2 \eta(1, 0, -1) \ ,\label{famdepcharge2}\ee 
The interfamily exponents of the Yukawa matrix associated with the
operator $L_i\overline{e}_jH_d$ follow:

\hskip 2cm
\begin{center}
\begin{tabular}{|c|c|c|c|}
\hline          
$ $ & ${\rm {\overline e}_1}$ & ${\overline e}_2$  & 
${\overline e}_3$  \\ \hline \hline   

${L}_1$ & $4$ & $5$ & $3$ \\

${L}_2$ & $1$ & $2$ & $0$ \\ 
                                                           
${L}_3$ & $1$ & $2$ & $0$ \\ \hline
\end{tabular} \ \ \ \ \ .
\end{center}
\vskip 0.3cm 

\noindent Its diagonalization yields the lepton interfamily hierarchy
\be{m_e \over m_{\tau}} \sim \lambda^4 \ , \ \ \ 
{m_{\mu} \over m_{\tau}} \sim \lambda^2 \ , \label{lephier}\ee
not inconsistent with phenomenology, as well as the contribution to
the lepton mixing matrix from the rotation of the left-handed lepton doublet:

\be
\left( \begin{array}{ccc}
1&\lambda^3&\lambda^3\\ \lambda^3&1&1\\  \lambda^3&1& 1
\end{array}  \right) \ .\label{lmix}\ee 
As $Y_F$ contains $B-L$, it is natural to introduce 
three families of right handed neutrinos ${\overline N}_i$. 
Before assigning them $Y_F$ charges, we note that certain predictions
associated with neutrino phenomenology are completely independent of the 
charges of the ${\overline N}$s. 
The neutrino mixing matrix, for example, is uniquely determined by the
charges of the MSSM fields \cite{RSGN}.  This is a result of its seesaw 
\cite{SEESAW} origin,
as can be seen via the following simple argument.
Since the right-handed neutrino Majorana mass matrix is symmetric,
it may be written 
$Y_{ij}^{(0)} = {\overline N}_i{\overline N}_j$, where ${\overline N}_i$ and 
${\overline N}_j$ are vectors depending only on the antineutrino charges.
The matrix coupling right-handed neutrinos to the standard model, on
the other hand, is written $Y_{ij}^{(\nu)}=H_uL_i{\overline N}_j$, where
$H_uL_i$ is a vector independent of the right-handed charges.  Taking $U^0$
to be the matrix that diagonalizes $Y^{[0]}$,

\be Y^{(0)} = U^0 D^0 (U^0)^T \ \ , \label{diag0}\ee
with $D^0$ a diagonal matrix, the effective neutrino mixing matrix
after the seesaw is given by:

\be \hat{Y}^{(\nu)} = - (Y^{(\nu)}U^0)(D^0)^{-1}(Y^{(\nu)}U^0)^T \ \ .
\label{seesaw}\ee
Because of the form of Eq. (\ref{seesaw}), a cancellation of
${\overline N}_i$ charges results, and one discovers 
that

\be \hat{Y}^{(\nu)}_{ij} = -H_u^2L_iL_j \ \ \ 
\label{seesaw2} \ \ \ .\ee 
The MNS neutrino mixing matrix \cite{MNS} therefore depends 
only on the mixing of the
$L_i$, a fact already noted in  
Refs. \cite{EIR,ILR}.  Thus, both the neutrino mass matrix and
the MNS mixing matrix appearing in the leptonic charged current are
determined
by the $L_i$ charges, and the MNS mixing matrix will be of the form
given in Eq. (\ref{lmix}).
\be
{\cal U}_{MNS}\sim \left( \begin{array}{ccc}
1&\lambda^3&\lambda^3\\ \lambda^3&1&1\\  \lambda^3&1& 1
\end{array}  \right) \ .\ee
This implies a small (order $\lambda^3$) mixing of the electron
neutrino 
with the $\mu$ and $\tau$ species, and mixing between the 
$\mu$ and $\tau$ neutrinos of
order one \cite{HRR}.  Remarkably enough, this mixing pattern is 
precisely the
one suggested by the non-adiabatic MSW \cite{MSW} explanation of the
solar neutrino deficit and by the oscillation interpretation of the
reported anomaly in atmospheric neutrino fluxes
\cite{superk,supk,soud}.  
It is important to stress that this mixing matrix is a generic prediction of
such models, and depends only on standard model charges already fixed
by phenomenology.  The neutrino masses, on the other hand, depend on
the origin of the intrafamily hierarchy.

\section{Intrafamily Hierarchy}


The intrafamily hierarchy in the quark sector suggests that
a family-independent symmetry is not the
end of the story.
Recall that the ratio of third
family quarks, $m_b/m_t$, is of order $\lambda^3$. Since both
$\cot\beta$ and the Yukawa entries conspire to produce this
suppression, there are two extreme possibilities.

$\bullet$ The first possibility
is that $Y_b$ and $Y_t$ are of the same order, with $\cot{\beta}$ 
responsible for the suppression.  
With a tree-level top quark mass, achieving $Y_b$ and $Y_t$ of the
same order requires that the $Y_F$ charge of the
$\mu$-term,
$H_uH_d$, be $Y_F^{[\mu]} = -6$.  But avoiding anomalies such as 
$Tr[YYY_F]$ and 
$Tr[SU(2)SU(2)Y_F]$ forces the $Y_F$ charge to be vector-like on 
the Higgs doublets, so that $Y_F^{[\mu]}=0$.
Hence $Y_b\sim Y_t$ requires $Y_F$ to be anomalous (The Green-Schwarz 
mechanism cannot be invoked since $Tr[SU(3)SU(3)Y_F]=0)$. Furthermore,
we shall soon see that a family-traceless $Y_F$ cannot reproduce 
neutrino phenomenology.

To proceed, we need to assign $Y_F$ charges to the 
right-handed neutrinos
$\overline{N}$.  Since $\eta$ is contained in $E_6$,
we give the $\overline{N}$ fields their $E_6$ value, $\eta=2$, which 
yields $Y_F(\overline{N}_i)=(-2,-1,3)$.  One 
obtains an $\overline{N}_i\overline{N}_j$ Majorana mass
matrix with family dependence

\hskip 2cm
\begin{center}
\begin{tabular}{|c|c|c|c|}
\hline          
$ $ & ${\overline N}_1$ & ${\overline N}_2$ & 
${\overline N}_3$  \\ \hline \hline   

${\overline N}_1$ & $4$ & $3$ & $SZ$ \\

${\overline N}_2$ & $3$ & $2$ & $SZ$ \\
                                                            
${\overline N}_3$ & $SZ$ & $SZ$ & $SZ$ \\ \hline
\end{tabular} \ \ \ ,
\end{center}
\vskip 0.3cm 
where `the SZ' stand for `supersymmetric zeros'
due to negative charges.
With a null row, this matrix has a zero eigenvalue, and the
third family neutrino drops out of the seesaw mechanism.
We are then left with two light
species of neutrinos, with masses
${v_u^2 \lambda^6 / M}$ and ${v_u^2 \lambda^{12} / M}$.
This situation is inconsistent with the combined set of 
atmospheric and solar neutrino data.  The predictions can be made
to fit any {\it one} experiment, however, but only if
$M$ is of order $10^{12}$~GeV,
suppressed by four orders of magnitude with respect to $M_{GUT}$.
There is no mechanism in our model to effect such
a suppression.  We conclude that the family-traceless, non-anomalous
$Y_F$ symmetry must be extended by adding a family-independent piece, 
hereafter called $X$.

$\bullet$ We turn now to the alternate possibility, 
$Y_b \sim \lambda^3Y_t$ and $\cot\beta$ of order 1,
where the suppression $\lambda^3$ comes from the family-independent
piece $X$. The total flavor symmetry is now

\be
Y_X \equiv X+Y_F  \ \ \ .
\ee
To consider the implications of anomalies involving our
family-independent
symmetry, we 
define the mixed anomaly coefficients of $Y_X$ with the 
Standard Model gauge
fields by

\be
C_{G_i} = Tr[G_iG_iY_X] \ . \label{ccoeff} \ee
The $C_{G_i}$ satisfy the following relations:

\be
C_Y + C_{\rm weak} - {8 \over 3}C_{\rm color} = 
6(X^{[e]}-X^{[d]})+2X^{[\mu]} \ \ , 
\label{anomrel1} \ee

\be
C_{\rm color}=3(X^{[u]}+X^{[d]})-3X^{[\mu]}
\label{anomrel2} \ \ , \ee
where $X^{[u,e,d,\mu]}$ are the $X$ charges of the operators
${\bf Q}_i{\bf \overline{u}}_jH_u$, $L_i\overline{e}_jH_d$, 
${\bf Q}_i{\bf \overline{d}}_jH_d$, and $H_uH_d$, respectively.  
It is precisely these charges $X^{[e]}$,
$X^{[d]}$, and $X^{[\mu]}$ that determine the 
intrafamily hierarchies $m_b / m_t$
and $m_{\tau} / m_b$.  Let us set
\be
{m_b \over m_t} \sim \cot\beta\lambda^{P_{bt}} \ , \ \ 
{m_{\tau} \over m_b} \sim \lambda^{P_{\tau b}} \ .
\ee
Then one finds that Eqs.~(\ref{anomrel1}) and
(\ref{anomrel2}) above
may be rewritten as
\be
C_Y + C_{\rm weak} - 2C_{\rm color} = -2(P_{bt}+3P_{\tau b}+6) \ ,
\label{powers}\ee
where we have used the fact that the top quark Yukawa coupling
appears at tree level, and therefore that $Y^{[u]}=0$.
The data suggest $P_{bt}=3$ and $P_{\tau b}=0$, which through
Eq. (\ref{powers}) tells us that our new symmetry $Y_X$
must be anomalous.  The only 
consistent way to build a model with such an anomalous $U(1)$ 
is the use of
the four dimensional version of the Green-Schwarz anomaly cancellation
mechanism \cite{GS}. We take the family-independent $X$ 
acting on the chiral fields to be a linear
combination of a universal piece and of the two $E_6$ charges, $V,V'$,
defined through
\be E_6 \supset SO(10) \times U(1)_{V'} \ \ \ ; \ \ \ 
SO(10) \supset SU(5) \times U(1)_V \ \ . \label{branch2}\ee
Across the Higgs doublets, the $X$ symmetry is taken to be
vector-like,
a necessary condition if the three $U(1)$ symmetries comprising
$Y_X$ are gauged separately \cite{ILR}.  These 
choices yield $X^{[d]}=X^{[e]}$, and
$X^{[u]}=X(LH_u\overline{N})=0$.
The Green-Schwarz structure has the added benefit of producing
the correct value of the Weinberg angle at cut-off \cite{IBAN,IR,BR}:
\be
\cot^2\theta_w={C_Y \over C_{\rm weak}}={5 \over 3} \ \ .
\ee

There still remains the non-zero anomaly $(YY_XY_X)$,  
which can be canceled  
by three families of standard model vector-like
representations ${\bf 5} + {\bf \overline{5}}$ of $SU(5)$.
With this addition made, the
remaining anomaly structure is consistent with the Green-Schwarz 
cancellation mechanism.  We get
\be
{m_b \over m_t}\sim\cot\beta\lambda^{-(C_{\rm color}+18)/3} \ , \ \ 
{m_{\tau} \over m_b}\sim1 \ ,
\ee
and agreement with the data is achieved for $C_{\rm color}=-27$.




We can now specify
the form of the matrices involving right-handed neutrinos:
\be Y^{(0)} \sim M\lambda^{-2X^{[\overline N]}-2} \left( \begin{array}{ccc}
\lambda^6&\lambda^5&\lambda\\ \lambda^5&\lambda^4&1\\
\lambda&1&\lambda ^{-4}
\end{array}  \right) \ \ \ \ \ ; \ \ \ \ \ 
Y^{(\nu)} \sim v_u\left( \begin{array}{ccc}
\lambda^8&\lambda^7&\lambda^3\\ \lambda^5&\lambda^4&1\\ \lambda^5&\lambda^4&1
\end{array}  \right) \ \ ,\label{nu1}\ee
where $Y^{(0)}$ is the $\overline{N}\overline{N}$ Majorana mass
matrix and $Y^{[\nu]}$   
the matrix coupling $L_i$ to ${\overline N}_j$.
Note that, to appear in the superpotential only as holomorphic quadratic mass
terms, the $X$-charge of the ${\overline N}$s must
be negative half odd integers.


After the seesaw, we have the actual neutrino mass matrix
\be \hat{Y}^{(\nu)} \sim {v_u^2 \lambda^{2X^{[\overline N]}+6} \over M}
\left( \begin{array}{ccc}
\lambda^6&\lambda^3&\lambda^3\\ \lambda^3&1&1\\ \lambda^3&1&1
\end{array}  \right) \ \ ,\label{nureal1}\ee
which produces light neutrinos with masses
\be
m_{\nu_e} \sim {v_u^2 \lambda^{2X^{[\overline N]}+12} \over M} \ \ ;
\ \ m_{\nu_\mu} \sim m_{\nu_\tau} \sim {v_u^2 
\lambda^{2X^{[\overline N]}+6} \over M} \ \ . 
\label{neutrinomass}
\ee
The mass splitting between $\nu_e$ and the other two neutrinos is 
$\Delta m_{\nu_{\mu}-\nu_e}^2\sim10^{-5}$eV$^2$, consistent with the 
non-adiabatic MSW solution to the solar neutrino
problem if $X^{[\overline{N}]}=-9/2$ and $M \sim M_{GUT}$.  
To check agreement with the
atmospheric neutrino data, we must know the mass splitting between
$\nu_{\mu}$ and $\nu_{\tau}$, but this can be
predicted only with a theory for the prefactors.  
Interestingly, as shown in
Ref. \cite{ILR},  
prefactors of order 1 produce 
$\Delta m_{\nu_{\tau}-\nu_{\mu}}^2\sim0.07$eV$^2$, 
so that the atmospheric data may be explained by the
same solution that accommodates the solar neutrino data without any fine
tuning.  
Moreover, this solution
requires $M$
to be of order $M_{GUT}$ as well, and drives the mixing angle to
maximal, in agreement with recent experimental
results \cite{superk,supk,soud}. 

As mentioned before, it is possible to gauge separately the three
symmetries that make up $Y_X$. The analysis proceeds much as
that above, but in this case
$X^{[d]}=-3$ instead of $X^{[d]}=-9$.  Also, as shown in \cite{ILR}, the extra
anomaly conditions fix all the charges of the $\mu$ term to zero, and
the analysis of the vacuum \cite{IL} in which all three symmetries are broken 
at the same scale favors  $X^{[\overline N]}=-3/2$.  
Remarkably, it is precisely
this charge assignment, corresponding to 
$X^{[\overline N]}=-9/2$ when the three symmetries are combined
into a single gauged symmetry $Y_X$, that leads to a fit 
of the neutrino data 
with $M~\sim~M_{GUT}$. 
We refer the interested reader to Ref. \cite{ILR}
for more details. 

\section{Acknowledgements}
JE would like to thank the Institute for Fundamental Theory at the 
University of Florida for its hospitality during the completion of
this work, and Kent State University for a summer grant in
support of the work.  NI and PR are supported in part by the 
United States Department of Energy under grant DE-FG02-97ER41029.


\begin{thebibliography}{Ref}

\bibitem{wolf} L. Wolfenstein, Phys. Rev. Lett. 51, 1945 (1983).
\bibitem{IR} L. Ib\'a\~nez and G. G. Ross, Phys. Lett. B332, 100 (1994).

\bibitem{BR} P. Bin\'etruy and P. Ramond, Phys. Lett. B350, 49 (1995).

\bibitem{superk} The Super-Kamiokande Collaboration, hep-ex/9805006 v2.
\bibitem{FN} C.~Froggatt and H.~B.~Nielsen Nucl. Phys. B147, 277 (1979).
\bibitem{RRR} P. Ramond, R.G. Roberts and G.G. Ross, Nucl. Phys. B406 (1993)
\bibitem{LENS} M. Leurer, Y. Nir, N. Seiberg, Nucl. Phys. B398,
319 (1993), Phys. Lett. B309, 337 (1993). 
\bibitem{RSGN} A. Rasin and J.P. Silva, Phys. Rev. D49, 20 (1994);
Y. Grossman and Y. Nir, Nucl. Phys. B448, 30 (1995).

\bibitem{SEESAW} M. Gell-Mann, P. Ramond, and R. Slansky in Sanibel
Talk, CALT-68-709, Feb 1979, and in {\it Supergravity} (North Holland,
Amsterdam 1979). T. Yanagida, in {\it Proceedings of the Workshop on
Unified Theory and Baryon Number of the Universe}, KEK, Japan, 1979.

\bibitem{MNS}Z. Maki, M. Nakagawa,and S. Sakata, Prog. Theo. 
Phys. 28, 247 (1962).

\bibitem{EIR} J. K. Elwood, N. Irges and P. Ramond,
Phys. Lett. B413, 322 (1997). 

\bibitem{ILR} N. Irges, S. Lavignac and P. Ramond, 
hep-ph/9802334, to be published in Phys. Rev. D.

\bibitem{HRR} J. A. Harvey, P. Ramond and D. B. Reiss, 
Nucl. Phys. B199, 223 (1982);  C. Carone and M. Sher,
Phys. Lett. B420, 83(1998); Carl H. Albright, K.S.~Babu, and
S.M.~Barr, contribution submitted to NEUTRINO 98 conference,
hep-ph/9805226; M.~Bando, T.~Kugo, and K.~Yoshioka,
Phys. Rev. Lett. 80, 3004 (1998).  

\bibitem{MSW} L. Wolfenstein, Phys. Rev. D17, 2369 (1978); S. Mikheyev and
A. Yu Smirnov, Nuovo Cim. 9C, 17 (1986).

\bibitem{supk} E. Kearns, talk at the ITP conference on Solar
Neutrinos:  News about SNUs, December 2-6 1997.

\bibitem{soud} S.M. Kasahara et al., Phys. Rev. D55, 5282 (1997).

\bibitem{GS} M. Green and J. Schwarz, Phys. Lett. B149, 117 (1984).

\bibitem{IBAN} L. Ib\'a\~nez, Phys. Lett. B303, 55 (1994).



\bibitem{IL}  N. Irges and S. Lavignac, Phys. Lett. B424, 293 (1998). 
G. Cleaver, M. Cvetic, J. R. Espinosa, L. Everett, P. Langacker,
CERN-TH-97-338, hep-th/9711178 .

\end{thebibliography}
\end{document}